\pdfoutput=1
\documentclass[amsmath,prd,aps,twocolumn,superscriptaddress,floatfix,nofootinbib]{revtex4}

\usepackage{amsfonts}
\usepackage{bm}
\usepackage{color}
\usepackage{graphicx}
\definecolor{darkgreen}{cmyk}{0.85,0.2,1.00,0.2}

\newcommand{\be}{\begin{equation}}
\newcommand{\ee}{\end{equation}}
\newcommand{\ba}{\begin{eqnarray}}
\newcommand{\ea}{\end{eqnarray}}
\newcommand{\nn}{\nonumber}

\newcommand\lsim{\mathrel{\rlap{\lower4pt\hbox{\hskip1pt$\sim$}}
        \raise1pt\hbox{$<$}}}
\newcommand\gsim{\mathrel{\rlap{\lower4pt\hbox{\hskip1pt$\sim$}}
        \raise1pt\hbox{$>$}}}

\def\n{{\hat{\bf n}}}
\def\r{{\bf r}}
\def\k{{\bf k}}
\def\l{{\bf l}}
\def\L{{\bf L}}
\def\C{\mathcal{C}}

\begin{document}

\title{Characterizing the Epoch of Reionization with the small-scale CMB: constraints on the optical depth and physical parameters}

\author{Simone Ferraro}
\affiliation{Berkeley Center for Cosmological Physics, University of California, Berkeley CA 94720, USA}
\affiliation{Miller Institute for Basic Research in Science, University of California, Berkeley CA 94720, USA}
\author{Kendrick M.~Smith}
\affiliation{Perimeter Institute for Theoretical Physics, Waterloo ON N2L 2Y5, Canada}

\date{\today}


\begin{abstract}
Patchy reionization leaves a number of imprints on the small-scale cosmic microwave background (CMB) temperature fluctuations, 
the largest of which is the kinematic Sunyaev-Zel'dovich (kSZ), the Doppler shift of CMB photons scattering off moving electrons in ionized bubbles.
It has long been known that in the CMB power spectrum, this imprint of reionization is largely degenerate with the kSZ signal
produced by late-time galaxies and clusters, thus limiting our ability to constrain reionization. Following Smith \& Ferraro (2017), 
 it is possible to isolate the reionization contribution in a model independent way,
by looking at the large scale modulation of the small scale CMB power spectrum.
In this paper we extend the formalism to use the full shape information of the small scale power spectrum (rather than just its broadband average), 
and argue that this is necessary to break the degeneracy between the optical depth $\tau$ and parameters setting the duration of reionization.
In particular, we show that the next generation of CMB experiments could achieve up to a factor of 3 improvement on the optical depth $\tau$ 
and at the same time, constrain the duration of reionization to $\sim 25 \%$. This can help tighten the constrains on neutrino masses, which will
be limited by our knowledge of $\tau$, and shed light on the physical processes responsible for reionization.

\end{abstract}


\maketitle

\section{Introduction}
Secondary anisotropies of the Cosmic Microwave Background (CMB) contain a wealth of information about the post-recombination Universe.
The largest small-scale temperature anisotropy that preserves the Black-Body spectrum of the CMB is the kinematic Sunyaev-Zel'dovich (kSZ) 
effect, that is the Doppler shift of CMB photons when scattering off moving electrons. The small-scale kSZ power spectrum receives roughly equal contributions from 
late-time galaxies and clusters, and from the epoch of patchy reionization  \cite{Ostriker:1986fua,Sunyaev:1980vz,Sunyaev:1972eq}. The late-time contribution informs us about the gas distribution in halos, the ionization
state of the intergalactic medium (IGM) and velocity fields at low redshift, while the reionization part has been shown to contain a wealth of information about the 
physical processes that drove it, as well as the redshift at which it happened \cite{Battaglia:2012im,McQuinn:2005ce,Park:2013mv, Alvarez:2015xzu, Zahn:2011vp}.  Since the two contributions have been shown to be comparable in amplitude and shape when 
looking at the power spectrum, a natural question is whether there is a robust way to disentangle them.

In recent work (Smith \& Ferraro, 2017 \cite{Smith:2016lnt}, henceforth SF17), 
we have shown that it is possible to isolate the reionization contribution in a model independent way, by looking at the long wavelength modulation of the locally-measured small scale kSZ power spectrum (which is a particular limit of the 4-point function). In brief, the small-scale kSZ power spectrum in a particular direction on the sky is modulated by the  realization of radial velocity along that line of sight.  Since the velocity field has a well-defined coherence length, it provides a ``standard ruler'' that allows the redshift of the source to be recovered.

Let's denote by $K(\n)$ the amplitude of the locally measured kSZ power spectrum in direction $\n$ and in some fixed high-$l$ band centered on $l_S$. In SF17, we have argued that $K(\n)$ fluctuates by order 10\% across the sky, and that its power spectrum $C^{KK}_L$, can robustly isolate the reionization contribution, without astrophysical uncertainties.

It is clear that there are two scales in the problem, namely the ``short'' wavelength mode $l_S \gtrsim 2000$ at which the kSZ power spectrum is obtained
 and the ``long'' mode $L \lesssim 300$ at which the modulation by velocity is measured.  We have shown that the $L$ dependence allows us to determine the redshift distribution of the source, with fluctuations at larger $L$ coming from higher redshift.  We refer the reader to SF17 for a detailed explanation of the method.

In this work we explore the $l_S$ dependence of the signal and show that, once the reionization component is isolated, it allows us to distinguish between different reionization models and constrain model parameters.  As a practical example, we will take the model of Battaglia et al \cite{Battaglia:2012id, Battaglia:2012im}, and show that we can distinguish between changes in optical depth $\tau$ and duration of reionization $\Delta z$.  Even in this simple parametrization, $\Delta z$ and $\tau$ are largely degenerate when not considering the $l_S$ dependence of the signal: For a fixed $l_S$, higher power can be obtained by either earlier reionization, so that the physical electron density contrast between neutral and ionized regions is higher, or by having a more extended reionization, so that CMB photons will encounter more ionized bubbles in their path and hence a larger anisotropy in the ionization field. This degeneracy is broken when measuring the signal as a function of $l_S$, since, as we shall see, a higher redshift of reionization (or equivalently $\tau$) leads to increase in kSZ power on all scales, while a longer reionization preferentially boosts the small scale power compared to larger scales. 

The remainder of the paper is organized as follows: In Section \ref{sec:formalism} we lay out the general formalism, and is Section \ref{sec:model} we describe a simple model based on simulations. In Section \ref{sec:results} we show numerical forecasts for future CMB experiments and discuss the dependence on survey design. We conclude in Section \ref{sec:conclusions}, discussing some of the challenges and future prospects.

\section{Formalism}
\label{sec:formalism}

We start by defining a set of $N_{\rm bins}$ filters $W_{S, i}(l)$ that are band-limited around multipole $l_{S,i} \gtrsim 2000$ and non-overlapping. For concreteness we take 
\be
    W_{S,i}(l)= 
\begin{cases}
    (C_l^{\rm kSZ})^{1/2}  / C_l^{\rm tot},& \text{if } l \in i\text{-th bin around } l_{S, i}\\
    0,              & \text{otherwise}
    \end{cases}
\ee
but we note that in the large number of bins limit, the $l$-weighting within a bin won't matter. 

Generalizing the approach of SF17, we consider $N_{\rm bins}$ fields $K_1(\n), \dots K_{N_{\rm bins}}(\n)$ representing the local small-scale power spectrum in direction $\n$, measured around multipoles $l_{S,i}$. More formally, given a filter $W_{S,i}(l)$, we apply it to the temperature map $T_{S, i}(\l) = W_{S, i}(l) T(\l)$ in harmonic space and obtain the $K_i$ by squaring $T_{S, i}$ in real space
\be
K_i(\n) = T_{S, i}^2(\n) 
\ee
We can then relate the redshift dependence of the sky-averaged $\bar{K}_i = \langle K_i(\n) \rangle$ to the kSZ power spectrum redshift source as
\be
\frac{d\bar K_i}{dz} = \int \frac{d^2\l}{(2\pi)^2} W_{S,i}(l)^2 \frac{dC_l^{\rm kSZ}}{dz}  \label{eq:dK_dz}
\ee
where $dC_l^{\rm kSZ}/dz$ is given explicitly by
\ba
\frac{dC_l^{\rm kSZ}}{dz} &=& 
  (T_{\rm CMB} \sigma_T n_{e,0})^2
   \frac{ (1+z)^4 }{\chi(z)^2 H(z)} 
   e^{-2\bar\tau(z)} 
\nn \\
&& \hspace{0.5cm} \times \, 
    \langle v_r(z)^2 \rangle 
   \, P_{ee}\!\left( \frac{l}{\chi(z)}, z \right)  \label{eq:dclksz_dz}
\ea
Here, $\langle v_r^2 \rangle = \langle v^2 \rangle/3$ is the mean squared radial velocity,
and $P_{ee}(k,z)$ is the spatial power spectrum of the electron overdensity field $\delta_e$.

In SF17 we have shown that an excellent approximation on large scales is to assume that the $K$ fields are modulated by the actual realization of the square of the line-of-sight velocity field,  $\eta(\n,z) = v_r(\n,z)^2 / \langle v_r(z)^2 \rangle$, so that
\be
K_i(\n) = \int dz\, \frac{d\bar{K}_i}{dz} \eta(\n,z)  \label{eq:K_eta}
\ee
which implies that in the Limber approximation,
\be
C_L^{K_i K_j} = \int dz \, \frac{H(z)}{\chi(z)^2} \left( \frac{d\bar K_i}{dz} \right) \left( \frac{d\bar K_j}{dz} \right)  P_\eta^\perp\!\left(\frac{L}{\chi(z)}\right)  \label{eq:clkk_ksz}
\ee
where $P_\eta^\perp$ is the power spectrum of $\eta$ with $\k$ perpendicular to the line of sight.
The power spectrum $C_L^{K_iK_j}$ in Equation~(\ref{eq:clkk_ksz}) represents clustering power due to correlations in the velocity field which sources kSZ anisotropy.
In addition, there is also ``noise'' power that would arise if the CMB were a Gaussian field, given by:
\be
N_L^{K_i K_i} = 2 \int \frac{d^2\l}{(2\pi)^2} W_{S, i}^2(\l) W_{S,i}^2(\L-\l) C_l^{\rm tot} C_{\L-\l}^{\rm tot}  \label{eq:nlkk}
\ee
where $N_L^{K_i K_j} = 0$ for $i \neq j$,
and $C_l^{\rm tot}$ is the total power spectrum of the temperature map,
including primary CMB, detector noise and residual foregrounds.

We build a Fisher matrix in terms of parameters $\pi_\alpha$ as
\be
F_{\alpha\beta} = \frac{f_{\rm sky}}{2} \sum_L (2L+1) \mbox{Tr} \left( \frac{\partial \C_L}{\partial \pi_\alpha} \C_L^{-1} \frac{\partial \C_L}{\partial\pi_\beta} \C_L^{-1} \right)  \label{eq:fisher}
\ee
where $(\C_L)_{ij} = C_L^{K_iK_j} + N_L^{K_iK_j}$ is an $N_{\rm bins}$-by-$N_{\rm bins}$ matrix,
and we have assumed that $K_i$ is a Gaussian field for purposes of computing the Fisher matrix.


The parameters that we consider are going to be a set of astrophysical parameters (such as timing or duration of reionization), and an additional $N_{\rm bins} (N_{\rm bins}+1)/2$ nuisance parameters, corresponding to an arbitrary constant (in $L$) shot-noise contribution to each $C_L^{K_i K_j}$, which will be marginalized over in all of our results. Marginalizing over an arbitrary constant in $C_L^{K_i K_j}$ will remove the contribution from any shot-noise components, such as any residual CIB or tSZ in the CMB maps. Moreover the contribution from CMB lensing has been shown to be reducible to white noise after the procedure described in SF17, and that is removed after marginalization.  As usual in the Fisher formalism, the marginalized parameter covariance is given by 
\be
{\rm Cov}(\pi_\alpha, \pi_\beta) = (F^{-1})_{\alpha \beta}
\ee

\section{Example: a simple reionization model}
\label{sec:model}
The formalism presented in the previous section can be applied to any parametrization of reionization. As an example, we shall consider the reionization modeling of Battaglia et al \cite{Battaglia:2012id, Battaglia:2012im}, where a semi-analytical model based on high-resolution radiation hydrodynamic simulations is applied to large volume $N$-body simulations in order to obtain the reionization kSZ field (among other quantities).  The simulation output is found to be well described by two phenomenological parameters, namely the mean redshift of reionization $\bar z$, and the reionization duration $\Delta z$.
Given an ionized fraction history $x_e(z)$, we define the mean optical depth $\bar \tau$ by:
\be
\bar \tau =  \int_0^{{z}_{\rm start}} \frac{dz}{H(z)} (1+z)^2 x_e(z) \ \sigma_T n_{p,0}  \label{eq:taubar}
\ee
where $n_{p, 0}$ is the proton comoving number density, $\sigma_T$ is the Thomson cross section and ${z}_{\rm start}$ is the redshift at the start of reionization.
We find that for models of reionization without extended components to high redshift and with mean redshift of reionization  $\bar{z}$ (defined such that $x_e(\bar{z}) = 0.5$), an excellent approximation is
\be
\bar \tau(\bar z) \approx \int_0^{\bar z} \frac{dz}{H(z)} (1+z)^2 \sigma_T n_{p,0}
\ee
While we use this approximation for simplicity, a given model of reionization will predict $x_e(z)$ which can be used self-consistently in Equation \ref{eq:taubar}, without any changes to our formalism.
There is one further subtlety here: Equation \ref{eq:taubar} neglects the correlation between fluctuations in ionization fraction and baryon density, which is usually present, as pointed out in \cite{Liu:2015txa}. The sign and amount of correlation of correlation depends on the mode of reionization and is still highly uncertain. Reference \cite{Liu:2015txa} estimated that neglecting it would typically correspond to an error on $\bar{\tau}$ of about $5\%$, depending on the model. However, measurements with future CMB and 21cm experiments will inform us on the particular characteristics of the reionization process, and allow us to further reduce modeling uncertainty. Moreover, the current uncertainty in cosmological parameters affects the relation by $\sim 1\%$ \cite{Liu:2015txa}, and will be negligible for future CMB experiments.

Here we take the parameters to be $\pi = \{\bar \tau, \Delta z, A_{\rm late}, \alpha_{\rm late} , A^{\rm s.n.}_{i, j}$\} where $\bar \tau = \bar \tau (\bar z)$ and $\Delta z$ are the parameters in the Battaglia et al simulations, $A_{\rm late}$ is the amplitude of the late-time kSZ contribution\footnote{Here defined in a model independent way as being the total contribution to $C_L^{K_i K_j}$ originating from $z < 6$.}, and $\alpha_{\rm late}$ is an arbitrary scale dependence of the form $l^{\alpha_{\rm late}}$, defined in Equation \ref{eq:kSZ_late}. As discussed in the previous Section, we also allow an arbitrary shot-noise component $A^{\rm s.n.}_{i, j}$ in each $C_L^{K_i K_j}$ as additional nuisance parameters that are marginalized over.

For the late-time contribution, we use a fit of $(d C_l^{\rm kSZ} / dz)_{\rm late}$ to the ``cooling and star formation'' (CSF) model of Shaw et al, normalizing the amplitude $A_{\rm late} = 1$ in the fiducial model. We further allow for an arbitrary scale dependence in the shape of the late-time profile of the form $l^{\alpha_{\rm late}}$ (with $\alpha_{\rm late} = 1$ in the fiducial model), so that 
\be
\left( \frac{d C_l^{\rm kSZ}}{d z} \right)_{\rm late} = A_{\rm late} \ l^{\alpha_{\rm late}} \left( \frac{d C_l^{\rm kSZ}}{d z} \right)_{\rm Shaw, CSF} \label{eq:kSZ_late}
\ee
The reason for allowing arbitrary $A_{\rm late}, \alpha_{\rm late}$ is that while different simulations agree on the fact that the late-time power spectrum should be featureless, they differ considerably on the predicted amplitude and slope. This is because they are both affected by complex sub-grid physical processes such as cooling, star formation and feedback, that are subject to large uncertainty.  Moreover, since the reionization contribution is also expected to be essentially featureless, this parametrization makes the two contributions completely degenerate when only looking at the power spectrum, while we will show that our method is able to efficiently separate them even in this case.  In this work, we'll assume no prior on $A_{\rm late}, \alpha_{\rm late}$, and note that better measurements of the late-time contribution along the lines of \cite{DeBernardis:2016pdv,Soergel:2016mce,Schaan:2015uaa,Hill:2016dta,Ferraro:2016ymw,Ade:2015lza,Hernandez-Monteagudo:2015cfa} may allow us to use tighter priors and further improve the reionization constraints.\

We will approximate the source redshift distribution of reionization kSZ as a Gaussian centered at $\bar{z}$, corresponding to a particular $\bar{\tau}$:
\be
\left( \frac{d C_l^{\rm kSZ}}{d z} \right)_{\rm rei}(z, l, \bar{\tau}, \Delta z) = (C_l^{\rm kSZ})_{\rm rei}(\bar{\tau}, \Delta z) \frac{e^{-(z-\bar{z})^2 / 2 \sigma_z^2}}{\sqrt{2 \pi \sigma_z^2}}
\label{eq:rei_der}
\ee
where we take the duration $\Delta z$ to be approximately the full-width at half maximum (FWHM) of the distribution, such that $\sigma_z \approx \Delta z / \sqrt{8 \ln 2}$.
The function $(C_l^{\rm kSZ})_{\rm rei}(\bar{\tau}, \Delta z)$ is a fit to the simulations by Battaglia et al \cite{Battaglia:2012im}, around the fiducial model with $\bar{z} = 8$ and $\Delta z$ = 1.2.

To calculate the Fisher matrix in Equation~\ref{eq:fisher}, we need to compute derivatives
of the form $(\partial C_L^{K_iK_j} / \partial\pi)$, where $\pi$ is one of the parameters
$\{ \bar\tau, \Delta z, A_{\rm late}, \alpha_{\rm late} \}$.
First we calculate parameter derivatives of $(dC_l^{\rm kSZ}/dz)$ directly from Equations~\ref{eq:kSZ_late} 
and~\ref{eq:rei_der}.  It is then straightforward to compute derivatives of $(d\bar K_i/dz)$ using Equation~\ref{eq:dK_dz},
then derivatives of $C_L^{K_iK_j}$ using Equation~\ref{eq:clkk_ksz}.

\section{Results}
\label{sec:results}
Here we show results for a next-generation CMB experiment such as CMB Stage 4 (CMB S4) type survey.  We shall assume a white detector noise with level $\Delta^2_T$ (usually quoted in $\mu$K-arcmin) and a Gaussian beam with full-width at half maximum (FWHM) $\theta_{\rm FWHM}$, such that the temperature map noise is given by
\be
N^{\rm det}_{l} =  \Delta^2_T e^{\theta^2_{\rm FWHM} l^2/(8 \ln2)} \,,
\ee

In Table \ref{tab:results} we shall consider the $l_S$ range 2000-7000, and show results for one $l_S$ bin as well as $N_{\rm bins} =20$ bins in $l_S$  on the same range. While the choice of 20 bins is arbitrary, we have checked that it saturates the amount of information present, so this should be taken to be the ``large number of bins'' limit.

\begin{table}[h!]
\begin{center}
  \begin{tabular}{| l | c | c |}
    \hline 
    \ $f_{\rm sky} = 0.7$ &\ $N_{\rm bins} = 1$ \ & \ $N_{\rm bins} = 20$ \ \\ \hline \hline
   \ $\sigma(\bar \tau)$ [marg.]  & 0.035 &  0.0028  \\ \hline
   \ $\sigma(\Delta z)$ [marg.]  & 3.92 & 0.32 \\ \hline
   \ $\sigma(\bar \tau)$ [unmarg.]  & 0.00022  & 0.00022  \\ \hline
   \ $\sigma(\Delta z)$ [unmarg.]  & 0.024 & 0.024 \\ \hline
  \end{tabular}
  \caption{Marginalized and unmarginalized constraints on $\bar{\tau}$ and $\Delta z$ with $f_{\rm sky} = 0.7$, beam $\theta_{\rm FWHM} = 1$ arcmin, map noise of $1\mu$K-arcmin. Both a single bin and 20 bins cases are considered.}
  \label{tab:results}
\end{center}
\end{table}

\begin{figure}[h]
\centerline{\includegraphics[width=9cm]{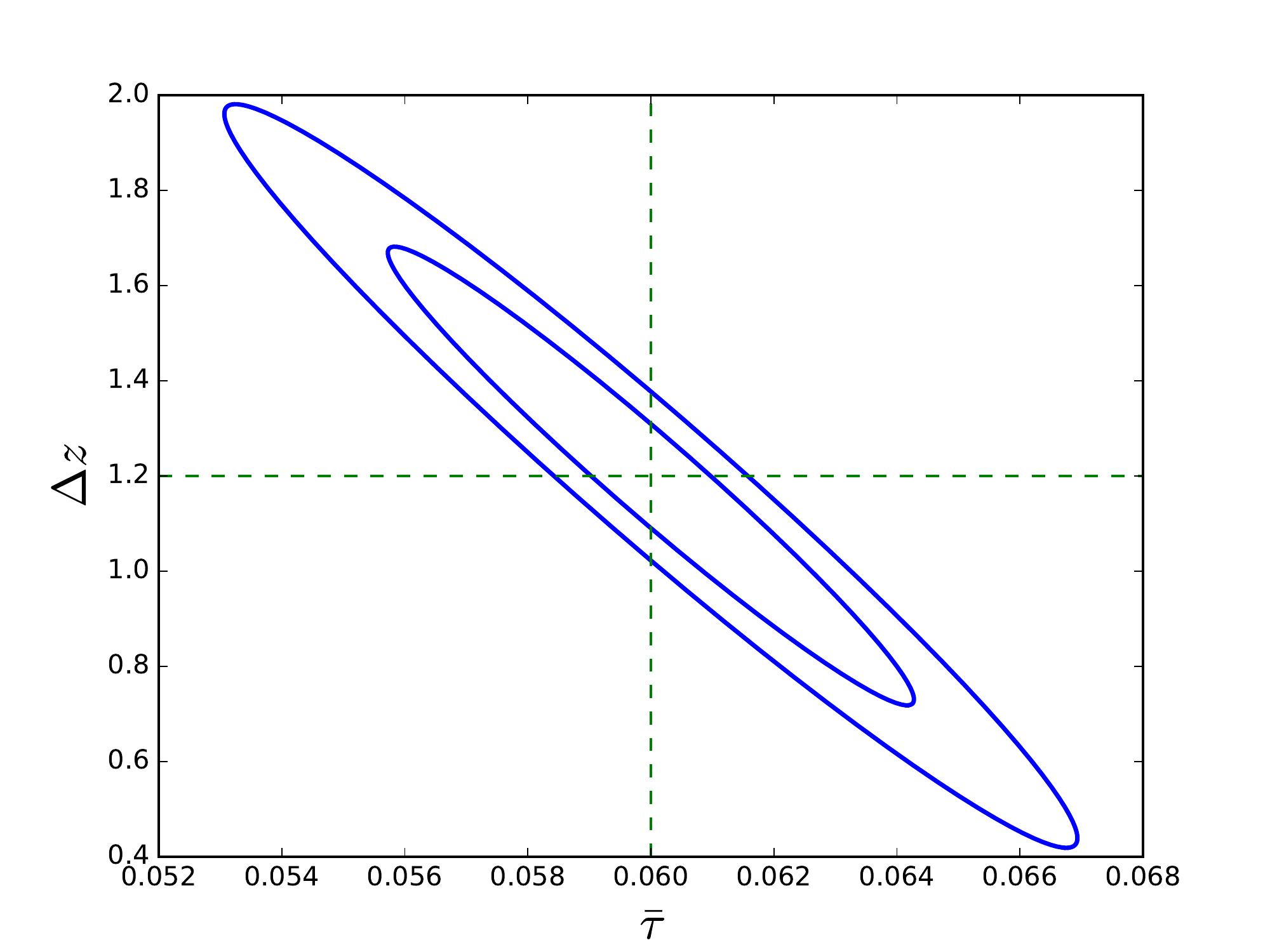}}
\caption{Marginalized constraints on the optical depth and reionization duration for a CMB S4-like configuration with $f_{\rm sky} = 0.7$, beam $\theta_{\rm FWHM} = 1$ arcmin, map noise of $1\mu$K-arcmin and $N_{\rm bins} =20$. The two ellipses correspond the 68\% and 95\% confidence levels.}
\label{fig:joint}
\end{figure}

\begin{figure}[ht]
\centering
\begin{tabular}{c}
  \includegraphics[width=9cm]{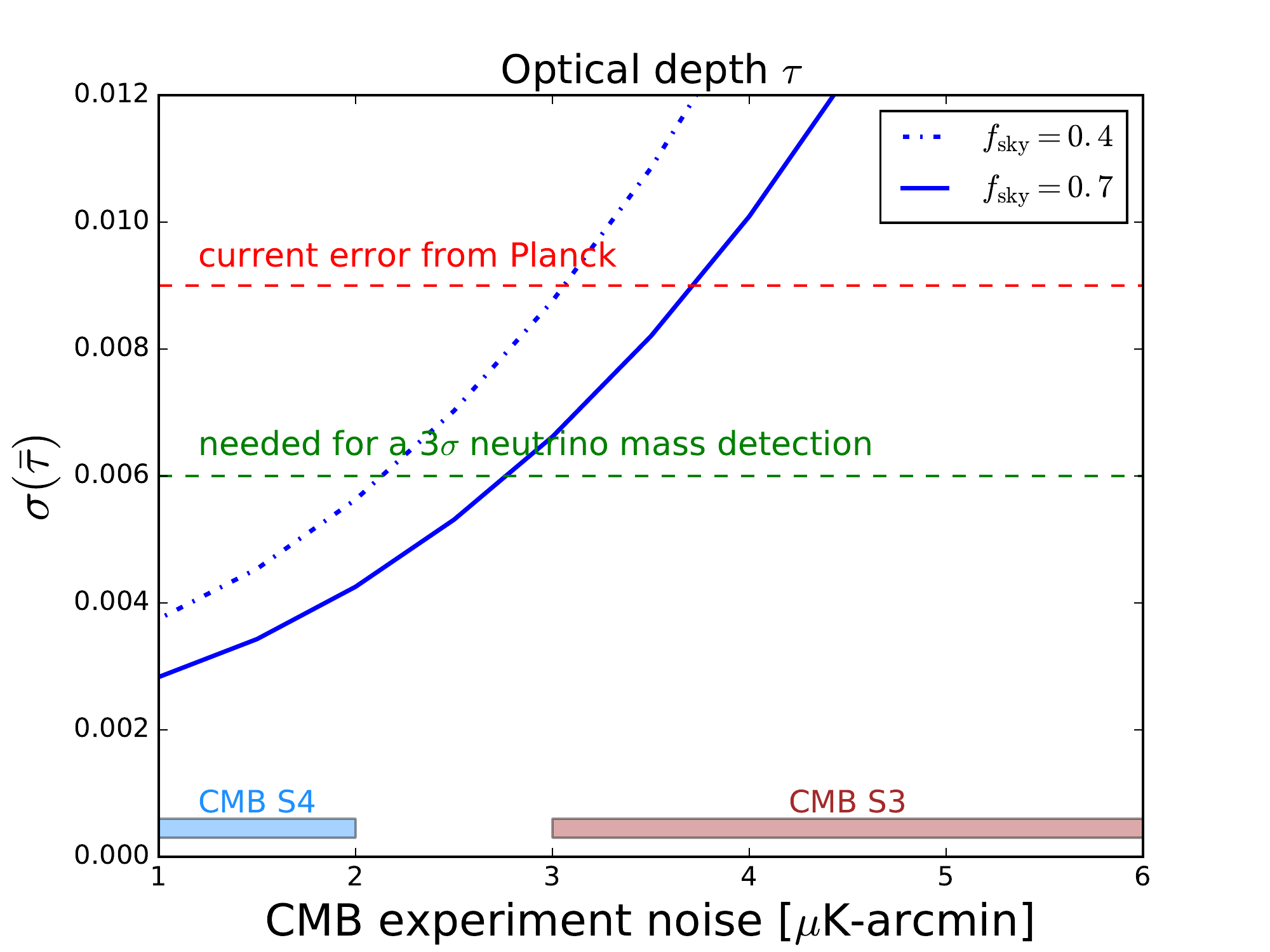} \\
  \includegraphics[width=9cm]{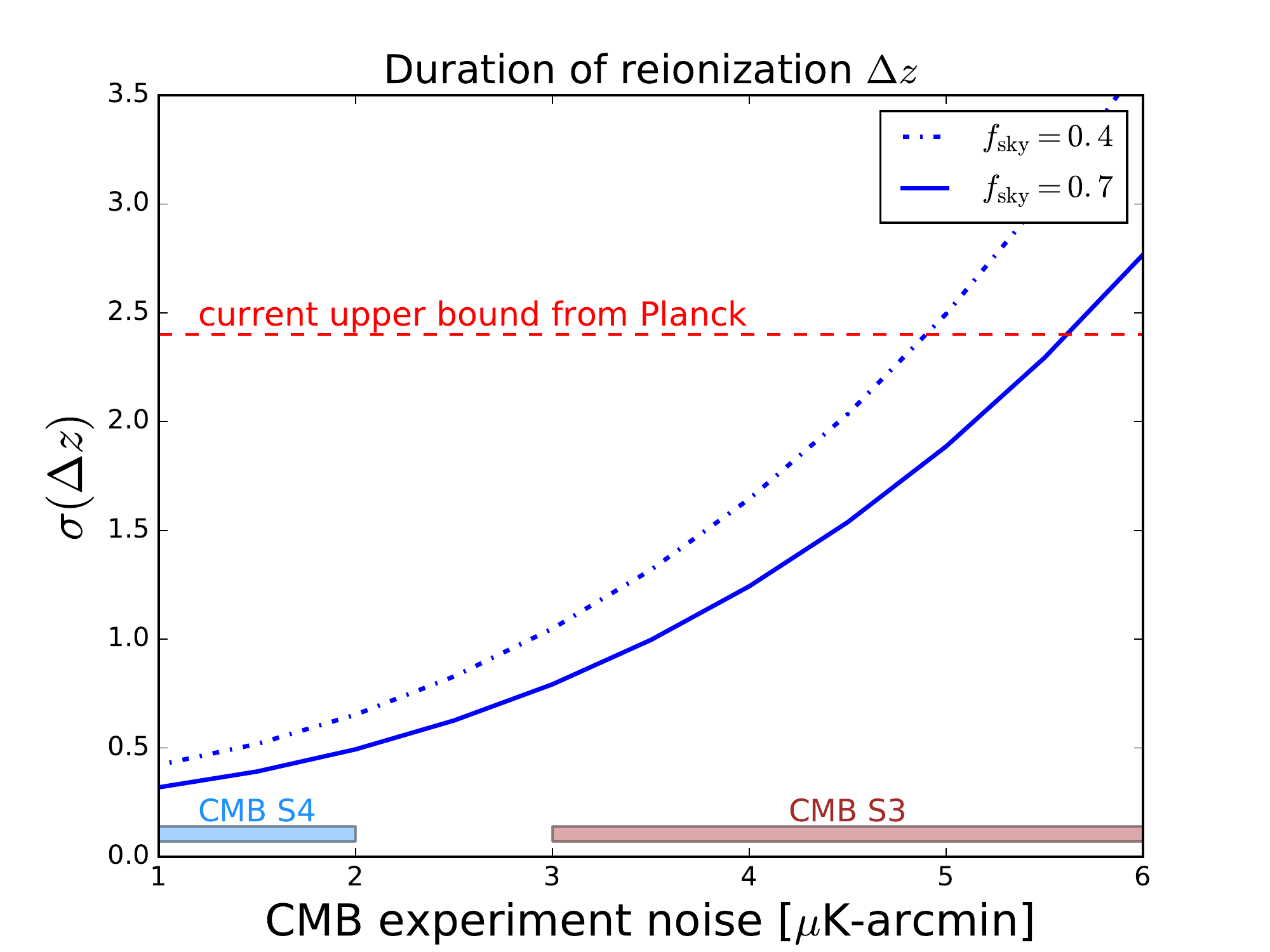} \\
\end{tabular}
\caption{Dependence of constraints on $\bar{\tau}$ (top) and $\Delta z$ (bottom), as a function of map noise, for a fixed beam $\theta_{\rm FWHM} = 1$ arcmin and $N_{\rm bins} =20$. Note that currently $\bar{\tau}$ has a $\sim 16 \%$ uncertainty from Planck large scale polarization. A $3 \sigma$ detection of minimum mass neutrinos in combination with DESI BAO would require an improvement in $\sigma(\bar{\tau})$ of $\sim 30\%$, within the reach of the next generation of CMB experiments.} 
\label{fig:noise}
\end{figure}

\begin{figure}[ht]
\centering
\begin{tabular}{c}
  \includegraphics[width=9cm]{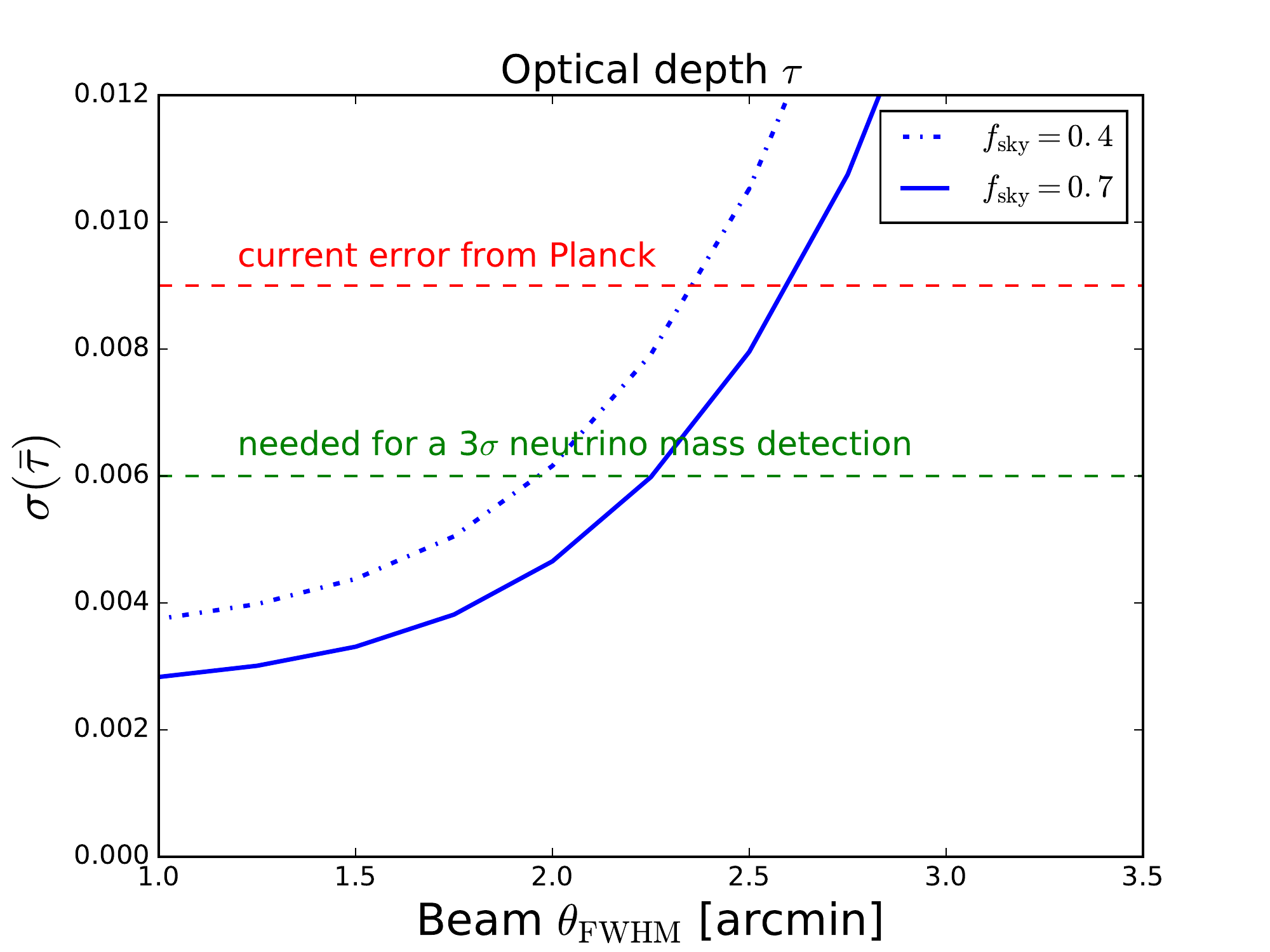} \\
  \includegraphics[width=9cm]{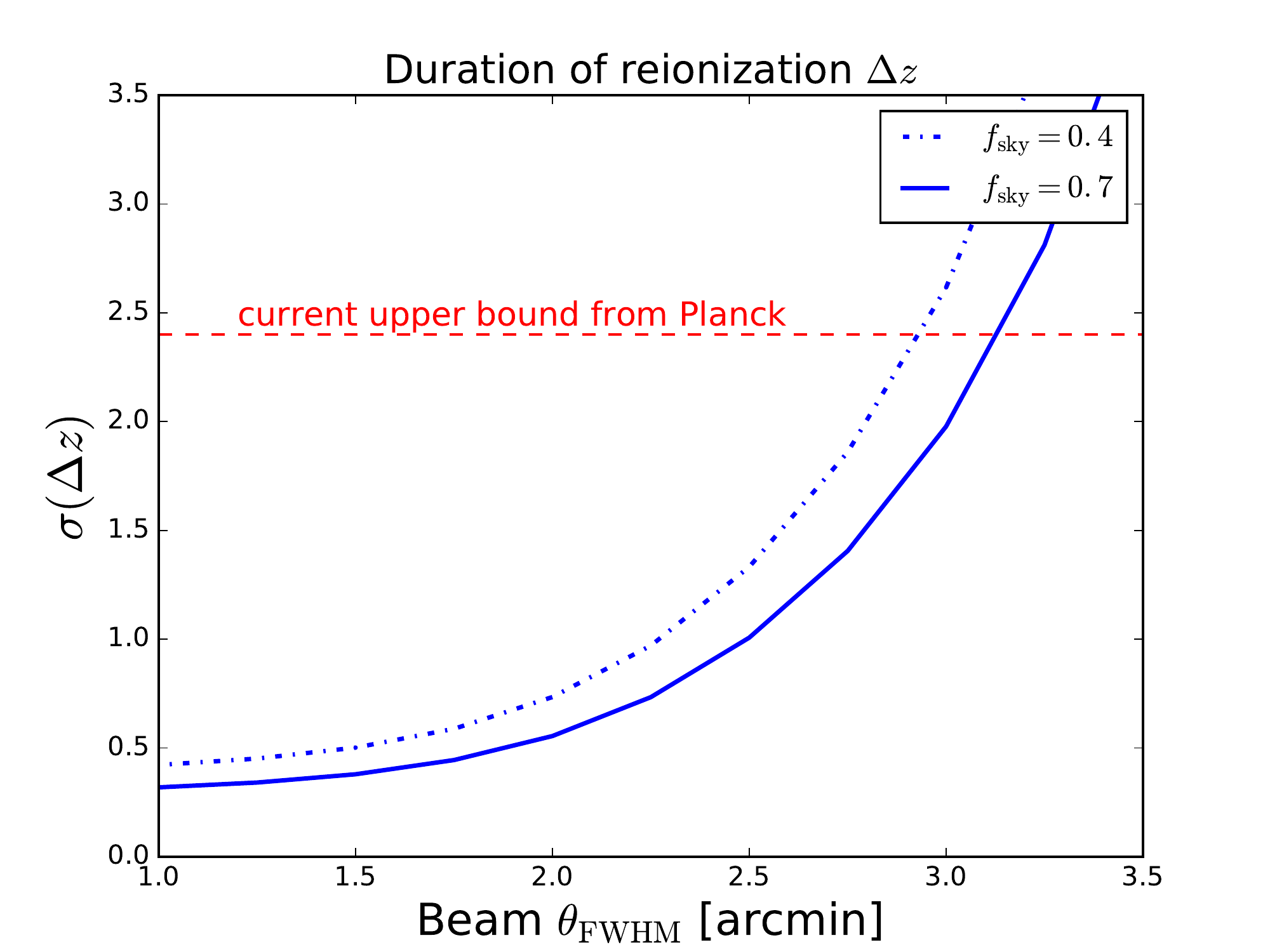} \\

  \end{tabular}
\caption{Dependence of constraints on $\bar{\tau}$ (top) and $\Delta z$ (bottom), as a function of beam $\theta_{\rm FWHM}$, for a fixed map noise = 1$\mu$K-arcmin and $N_{\rm bins} =20$. Note that the currently $\bar{\tau}$ has a $\sim 16 \%$ error from Planck, while Planck provides an upper bound on $\Delta z$ from the kSZ power spectrum.} 
\label{fig:beam}
\end{figure}

First we note that even in the case of a single bin, the reionization duration and optical depth are not perfectly degenerate as might naively be expected. This is because the $L$ shape of $C_L^{KK}$ itself (and in particular the position of the peak) contains information about the redshift of reionization: an earlier reionization will move the peak to higher $L$ because a fixed comoving scale (the velocity coherence length in this case), will subtend a smaller angle in the sky. Our formalism optimally combines the information in both the $L$ and $l_S$ dependence to constrain parameters.

When considering the case with $N_{\rm bins} = 20$, corresponding to 20 distinct $K_i$ fields, we see that there is considerable additional information in the $l_S$ dependence: marginalized errors on $\bar \tau$ and $\Delta z$ are reduced by a factor of over 12 compared to the single bin case.

Table \ref{tab:results} summarizes our main results. When marginalizing over all the nuisance parameters\footnote{These include one between $\bar \tau$ or $\Delta z$, as well as the late-time kSZ amplitude $A_{\rm late}$ and scale dependence $\alpha_{\rm late}$, and an arbitrary shot-noise component in each of the $C_L^{K_i K_j}$.}, we find that in our simple model we obtain $\sigma(\bar \tau) = 0.0028$, for a future experiment with a 1 arcmin beam and 1 $\mu$K-arcmin noise on 70\% of the sky.
This corresponds to a $\approx 5\%$ measurement of the optical depth, an improvement by a factor of $\approx 3$ over the current Planck measurement from the large scale CMB polarization \cite{Aghanim:2016yuo}. At the same time, we can constrain the duration of reionization to $\sim 25\%$. Figure \ref{fig:joint} shows the joint constraints between $\bar{\tau}$ and $\Delta z$, illustrating the large degeneracy between the two, even in the $N_{\rm bins} = 20$ case. This is also clear in Table \ref{tab:results}, where we show that marginalization over the other parameter (as well as the additional nuisance parameters), degrades constraints by a factor of $\sim 10$ for $N_{\rm bins} = 20$ and by $\sim 100$ for $N_{\rm bins} = 1$.

Next, we study the dependence on experimental configuration for the $N_{\rm bins} = 20$ case, and results are shown in Figures \ref{fig:noise} and \ref{fig:beam}.  We find an especially steep dependence on resolution, with constraints on $\bar \tau$ or $\Delta z$ improving by a factor of $\approx 5$ and 6 when the beam size is reduced from 3 arcmin to 1.5 arcmin. An experimental configuration with beam larger than $\sim 2.5 - 3$ arcmin provides no improvement on reionization compared to the current bounds from Planck.  The dependence on map noise is shallower, with improvement over Planck expected for noise $\lesssim 4 \mu$K-arcmin.

\section{Discussion}
\label{sec:conclusions}
\subsection{Optical Depth and Neutrino Masses}
It is well known that cosmological observations are sensitive to the sum of neutrino masses $\sum m_\nu$ (see for example \cite{Lesgourgues:2014zoa}). Physically, neutrinos suppress the growth of structure on scales smaller than their free-streaming length, and a measurement of their mass is obtained by comparing the primordial amplitude $A_s$ at CMB to some tracer of the amplitude of fluctuations at late times, such as CMB lensing, or galaxy lensing or clustering. However, the CMB only measures the combination $A_s e^{-2\tau}$ (except on the very largest scales), and therefore knowledge of $\tau$ is essential for a neutrino mass measurement.  In fact, forecasts show that for a future generation CMB experiment such CMB S4 (both primary power spectrum and CMB lensing), the uncertainty on $\tau$ will be the limiting factor on the measurement of $\sum m_\nu$ \cite{Abazajian:2016yjj,Allison:2015qca}. The Planck mission has measured $\tau$ from the very low $l$ polarization data, with $\sigma(\tau) \approx 0.009$ \cite{Aghanim:2016yuo}. Since CMB S4 will likely only access scales with $l \gtrsim 30$, it will not improve the existing constraints, and will have to rely on previous measurements. It can be shown that CMB S4, together with an expansion probe such as Baryon Acoustic Oscillations (BAO) from DESI and the Planck $\tau$ prior, will obtain $\sum m_\nu \approx 30$ meV \cite{Abazajian:2016yjj}, not sufficient for a reliable detection of the minimum neutrino mass implied by oscillation experiments $(\sum m_\nu)_{\rm min} = 58$ meV. Similar results hold when using LSST as a low-redshift tracer instead, since the $\tau$ degeneracy is the limiting factor, not the measurement of the late time amplitude.  While the exact number depends on the configuration of the CMB S4 experiment, a $3 \sigma$ `detection' of minimal mass neutrinos requires $\sigma(\tau) \approx 0.006$, or roughly 40\% better than the current constraints (see Figure 16 of \cite{Abazajian:2016yjj}).  Similarly, a $4 \sigma$ detection would need $\sigma(\tau) \lesssim 0.003$. Dedicated space missions have been proposed \cite{Matsumura:2016sri,Bouchet:2015arn,Matsumura:2013aja}, and there are ground based experiments aiming to measure large-scale polarization from the ground \cite{Harrington:2016jrz}.

\subsection{The physics of reionization and complementary probes}
The measurements proposed here are highly complementary to other probes of the epoch of reionization. Observations of high-redshift Ly$\alpha$ and Ly$\beta$ forest suggest that reionization was essentially complete by $z \approx 6$ \cite{McGreer:2014qwa}.  Measurements of the redshifted 21cm emission (and other lines) by experiments such as HERA  \cite{DeBoer:2016tnn} will be highly complementary, since CMB will have an order of magnitude better angular resolution ($\sim 1$ arcmin vs $\sim 10$ arcmin), while 21cm has higher redshift resolution. Moreover, the kSZ signal is independent of temperature and therefore less affected by modeling uncertainties.  Even with a detection of reionization by the next generation of 21cm experiments, model parameters can still be highly degenerate, since they have been shown to have a similar physical effect \cite{Kern:2017ccn}. The higher resolution of CMB experiments can help break degeneracies, and a joint analysis is expected to be highly beneficial.  Moreover, cross correlating the kSZ field (or the $K$ field defined in this work) with higher moments of the 21cm temperature can reduce the impact of foregrounds that severely affect the power spectrum from intensity mapping experiments.
\subsection{Modeling Uncertainties}
One important uncertainty in the above discussion is the presence of foregrounds in CMB temperature maps.  A multi-frequency analysis can mitigate their impact and all of the noise levels in the paper should be intended as the temperature map noise after component separation.  In a realistic situation, the residual noise might not be perfectly white, and there may be residual foreground non-Gaussianity.  Detailed modeling of the effect of foregrounds and component separation, which requires realistic multi-components simulations, is the subject of ongoing work.  

Another important question concerns the model dependence of the constraints on reionization. Firstly, we note that while we have used a simple model as an example of the use of our formalism, any parametrization of reionization can be treated in the same way.

We should also note that the separation of the signal into low-redshift and reionization only relies on knowing the large-scale velocity power spectrum, which is well described by linear theory. Therefore, a measurement of the amplitude and shape of the reionization kSZ power spectrum can be obtained with very little astrophysical uncertainty (modulo foregrounds), and extracting this being the goal of the new technique introduced in SF17 and extended in this work.

In this paper, we studied a specific reionization model due to Battaglia et al~\cite{Battaglia:2012im}
and found that the kSZ-derived statistic $C_L^{K_iK_j}$ was sufficient to fully break degeneracies
and determine all parameters in the model.  One may wonder whether this conclusion is specific
to the Battaglia et al model, or applies generally to
parameterized models of reionization.  We defer a complete study to future work, but
in the next few paragraphs, we will give a heuristic argument which suggests that it should be 
a fairly generic conclusion under certain assumptions which we will state explicitly.

First, we note that in most reionization models, the ionization fraction $x_e(\r,z)$ is
approximately either 0 or 1 everywhere, since reionization fronts are expected to be very sharp compared to the size of the bubbles.
Therefore, we have
\be
\bar x_e(z) \approx \langle x_e(\r,z)^2 \rangle  \label{eq:xe_xe2}
\ee
since $x_e \approx x_e^2$.
During reionization it is also a good approximation to neglect cosmological density
fluctuations and attribute all variations in electron density to variations in ionization
fraction, i.e.~we have $\delta_e(\r,z) \approx x_e(\r,z) - \bar x_e(z)$.
We take the expectation value of the square on both sides, and simplify as follows:
\be
\langle \delta_e(\r,z)^2 \rangle = \int_0^\infty dk \, \frac{k^2}{2\pi^2} P_{ee}(k,z)
\ee
\ba
\Big\langle \Big( x_e(\r,z)^2 - \bar x_e(z) \Big)^2 \Big\rangle &=& \langle x_e(\r,z)^2 \rangle - \bar x_e(z)^2 \nn \\
  & \approx & \bar x_e(z) (1 - \bar x_e(z))
\ea
where we have used Eq.~(\ref{eq:xe_xe2}) in the last line.
Putting this together, we obtain the following approximate relation between the mean
ionization fraction $\bar x_e(z)$ and the electron power spectrum $P_{ee}(k,z)$:
\be
\bar x_e(z) (1 - \bar x_e(z)) \approx \int_0^\infty dk \, \frac{k^2}{2\pi^2} P_{ee}(k,z)  \label{eq:xe_pe}
\ee
Next we note that our observable $C_L^{K_i K_j}$ is a function of two variables: the large 
scale $L \lsim 300$ and the CMB scale $l_S \approx 4000$ which is implicit in the $i,j$ indices.  
By Eqs.~(\ref{eq:dK_dz}),~(\ref{eq:dclksz_dz}), the observable $C_L^{K_i K_j}$ is linearly sourced by another function of 
two variables: the power spectrum $P_{ee}(k,z)$.
Since both the observable $C_L^{K_iK_j}$ and the source $P_{ee}(k,z)$ are functions of two variables,
then there are enough degrees of freedom to solve for $P_{ee}(k,z)$, when $C_L^{K_iK_j}$ is observed.
Of course the effective redshift resolution in $P_{ee}(k,z)$ will be limited, due to correlations which
arise between nearby redshifts.
However, if the reionization model only allows slowly varying time dependence, we should have
have enough constraining power to fit for all model parameters which determine $P_{ee}(k,z)$.

Now we point out that a measurement of $P_{ee}(k,z)$ is sufficient to determine $\bar\tau$.
By Eq.~(\ref{eq:xe_xe2}) we see that $P_{ee}(k,z)$ determines the quantity $\bar x_e(z) (1 - \bar x_e(z))$ as a function of $z$.
Assuming that $\bar x_e(z)$ is an increasing function of $z$, this suffices to determine $\bar x_e(z)$.
Finally, $\bar x_e(z)$ determines $\bar\tau$ by Eq.~(\ref{eq:taubar}).

This heuristic argument can be summarized by saying that the same underlying source function
$P_{ee}(k,z)$ determines both $\bar\tau$ and the kSZ observable $C_L^{K_iK_j}$, and that the
latter has enough degrees of freedom (being a function of two variables) to solve for $P_{ee}(k,z)$
and infer $\bar\tau$.

Let us now consider the assumptions in the argument.
First, as previously noted, the reconstruction of $P_{ee}(k,z)$ will have large off-diagonal correlations 
in redshift, and our argument is likely to break down for reionization models which allow very rapid redshift dependence.
Second, we have assumed that $x_e$ is 0 or 1 everywhere, i.e.~reionization occurs in sharply defined bubbles.
Third, the $k$-integral in Eq.~(\ref{eq:xe_pe}) formally runs from $k=0$ to $k=\infty$, whereas the kSZ observations
only cover a finite range of scales near $k \sim 0.7$ $h^{-1}$ Mpc (corresponding to $l_S \sim 4000$ in
the CMB).  If the reionization model contains enough parameters that these scales do not suffice to determine
the integral in Eq.~(\ref{eq:xe_pe}), then our argument may break down.  For example, Ref.~\cite{Park:2013mv} considers
models with a low level of reionization out to very high redshift, where (crucially) the reionization bubbles
are very small, and hidden below the CMB beam scale.
Fortunately, measurements of the shape of the low-$l$ CMB polarization power spectrum 
have the potential to test this scenario \cite{Miranda:2016trf,Heinrich:2016ojb}.

The argument we have given above is heuristic, and more quantitative work is necessary to test the assumptions made.  Nevertheless, we find it useful for explaining qualitative properties of the kSZ statistic $C_L^{K_iK_j}$ at an intuitive level, and point out that the results may be fairly insensitive to the detailed modeling of reionization.

In summary, the kSZ power spectrum from reionization can be isolated robustly from the late-time kSZ contribution, and it is sensitive enough to $\tau$ and other parameters to have the \textit{statistical potential} for great improvement over the current uncertainties. The measurement will be robust to the extent that reionization doesn't have long tails in redshift and the model used in the analysis encompasses the truth. While no single measurement can guarantee that, agreement of model predictions for multiple observables (such as 21cm or other line intensity mapping experiments) will likely be the most robust way to give confidence on the correctness of the model. The importance of $\tau$ for cosmology and the statistical power of the kSZ effect, motivates further research into the interpretation of the signal.

\section{Conclusions}
Building on our previous work in SF17~\cite{Smith:2016lnt}, we have introduced a new formalism to isolate and characterize patchy reionization in the high-$l$ CMB. By optimally combining all of the information in the kSZ 4-point function, we showed that the reionization and late-time parts of the signal can be isolated, even in the case when they are completely degenerate in the power spectrum. To show this, we have allowed for an arbitrary amplitude and scale dependence of an otherwise featureless late-time kSZ power spectrum, and marginalized over those, as well a large number of other nuisance parameters aimed at absorbing the effect of foregrounds and CMB lensing.

Using a simple model of reionization, we have found that the next generation of ground based CMB experiments such as CMB S4 will have the statistical power to significantly improve on the current measurements of the optical depth $\tau$ and the duration of reionization $\Delta z$.  While some uncertainty over the correct modeling of reionization and foregrounds persist, the great statistical power shown makes further study worthwhile. Moreover, a number of integral constraints on the reionization power spectrum may make the observables less sensitive to the particular model adopted in the analysis.

Measuring $\tau$ from the next generation of ground-based experiments appears challenging, because of the difficulty in measuring the very large scale polarization signal. Therefore, short of a dedicated space mission, the high-$l$ kSZ signal that we have discussed might be one of the most promising ways to break the optical depth degeneracy and measure neutrino masses from cosmology.

\acknowledgements
We are grateful to Nick Battaglia for providing the kSZ power spectra from the simulations in \cite{Battaglia:2012im}, and Marcelo Alvarez, Adrian Liu, Matt McQuinn, Emmanuel Schaan, Uro${\rm \check{s}}$ Seljak, Blake Sherwin, David Spergel and Martin White for very useful discussion.
SF was supported by the Miller Institute at the University of California, Berkeley.
KMS was supported by an NSERC Discovery Grant and an Ontario Early Researcher Award.
Research at Perimeter Institute is supported by the Government of Canada
through Industry Canada and by the Province of Ontario through the Ministry of Research \& Innovation.

\bibliographystyle{prsty}
\bibliographystyle{h-physrev}
\bibliography{tau_ksz}

\end{document}